\documentclass[onecolumn]{revtex4}
\usepackage{amsmath,amssymb,graphics,epsfig,subfigure}
\usepackage{color}
\usepackage[colorlinks,linkcolor=red,anchorcolor=red,citecolor=green]{hyperref}
\usepackage{setspace}
\usepackage{booktabs}
\usepackage{float}
\begin{document}
	\thispagestyle{empty}
\begin{center}
	\title{The Kramers escape rate of phase transitions for the 6-dimensional Gauss-Bonnet AdS black hole with triple phases}
	
	%\date{\today}
	\author{Chen Ma$^{1}$, Pan-Pan Zhang$^{1}$, Bin Wu$^{1,2,3}$, Zhen-Ming Xu$^{1,2,3,}$\footnote{E-mail: zmxu@nwu.edu.cn}
		\vspace{6pt}\\}
	
	\affiliation{$^{1}$School of Physics, Northwest University, Xi'an 710127, China\\
		$^{2}$Shaanxi Key Laboratory for Theoretical Physics Frontiers, Xi'an 710127, China\\
		$^{3}$Peng Huanwu Center for Fundamental Theory, Xi'an 710127, China}
	
	\begin{abstract}
In this study, we obtain specific picture of the phase transitions for the 6-dimensional Gauss-Bonnet Anti-de Sitter (AdS) black hole with triple phases, using the generalized free energy we constructed and Kramers escape rate in stochastic motion. There are six possible phase transition processes between the three different stable states (small, medium, and large black hole states). During these phase transitions, there are two key temperatures. One is the temperature at which the medium black hole state emerges, and the other is the temperature at which the small black hole state annihilates. Meanwhile, two dynamic equilibrium processes are formed. One is a dynamic equilibrium of the transition from the medium black hole state to the large black hole state and the transition from the small black hole state to the medium black hole state. The other is a dynamic equilibrium of the transition from the small black hole state to the medium black hole state and the transition from the medium black hole state to the small black hole state.
	\end{abstract}

\maketitle
\end{center}

\section{Introduction}
Black hole thermodynamics, an effective theory for understanding the essential properties of gravity, has received widespread attention and research. It originate from the pioneering works~\cite{Hawking1974,Bekenstein1973,Hawking1976} of Hawking and Bekenstein, who insisted that black holes have temperature and entropy. The high similarity between the four laws of black hole mechanics and thermodynamics further pushes the study of black hole physics into the field of thermodynamics~\cite{Bardeen1973}. With the subsequent introduction of extended phase space~\cite{Kastor2009,Kubiznak2012}, holographic extended phase space~\cite{Visser2022,Cong2022}, and constrained phase space~\cite{Gao2022}, the thermodynamic phase transition of black holes has become an indispensable and important research topic in black hole physics. The Hawking-Page transition~\cite{Hawking1983} demonstrates the transition between the AdS black holes and thermal radiation states. The negative cosmological constant is considered as the black hole pressure~\cite{Kastor2009}, resulting that van der Waals-like phase transitions exist in the charged AdS black holes~\cite{Kubiznak2012}. Black hole molecules, Ruppeiner thermodynamic geometry and thermodynamic topology are used to analyse the possible microscopic mechanism of black hole phase transition~\cite{Wei2015,Miao2018,Wei2019,Xu2020,Ghosh2020,Wei2022a,Wei2024,Xiao2024}.

Over the past 50 years of development, the study of black hole thermodynamics has become increasingly extensive, gradually beginning to explore from the surface to the inside, and gaining a deeper understanding of the underlying physical mechanisms behind the thermodynamic critical phenomena of black holes. Previous studies have mostly focused on using equilibrium thermodynamic theory to quantitatively characterize the phase transition and critical behavior of black hole thermodynamic systems. But when we ask how black hole phase transitions occur, we need to involve relevant theories of non-equilibrium statistics, which also remain a challenging issue in traditional thermodynamic theory.

At present, some attempts have been made to investigate the dynamic processes of thermodynamic phase transitions in black holes. Some studies attempt to obtain the underlying structure of phase transition by establishing relativistic stochastic statistical physics~\cite{Cai2023a,Cai2023b}. Some studies attempt to use the first passage time in stochastic motion to obtain the dynamic information of phase transition~\cite{Li2020,Wei2021,Cai2021}. There are also some works attempt to use the generalized free energy to obtain the rate behavior of phase transition~\cite{Xu2023,Wang2024a,Xu2024a,Liu2023,Du2023,Sadeghi2024a,Sadeghi2024b}. In this study, we use the previously proposed generalized Maxwell equal area law to obtain some procedural descriptions of three-phases phase transitions for the 6-dimensional Gauss-Bonnet AdS black hole.

In terms of the three-phases phase transition of black holes, previous studies have focused on analyzing the triple point and corresponding critical exponents, but lacks a description of the transition process between the three different stable phases. Which process takes priority in these phase transition processes? Are these different processes starting simultaneously? End at the same time? Or in other ways? The answers to these questions help us gain a clearer understanding of the characteristics of the three-phases phase transitions in black holes. We attempt to answer these questions by using generalized free energy and Kramers escape rate in Brownian motion to obtain process information of different phase transitions, in order to provide a clear picture of three-phases phase transitions. We take a 6-dimensional Gauss-Bonnet AdS black hole as an example for discussion, which is a typical thermodynamic system of black holes with three-phases phase transitions. By calculating the six possible phase transition processes between these three different stable states, a clear phase transition picture of this black hole thermodynamic system is provided. Next, we will start building our specific approach and describing the physical results obtained.

\section{Generalized free energy and Kramers escape rate}
\begin{figure}[htb]
	\centering
	\subfigure[at the constant temperature]{
		\includegraphics[width=65 mm]{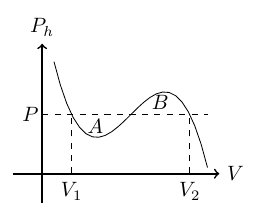} }
	\subfigure[at the constant pressure]{
		\includegraphics[width=65 mm]{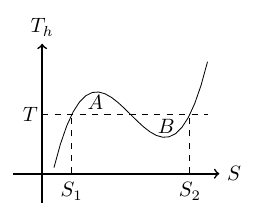} }
	\caption{The Maxwell equal area law of the thermodynamic system of black holes.}\label{fig1}	
\end{figure}

In the extended phase space of the black hole thermodynamics, the precise analogy between the small-large black hole phase transition and the gas-liquid phase transition of van der Waals fluid has benn realized~\cite{Kubiznak2012}.
Therefore, the phase transition of the black hole can be determined by the Maxwell equal area law, which is a very important tool in calculating the gas-liquid phase transition.  For the thermodynamic system of black holes, we can obtain its equation of state, i.e., the relationship $P_h=P_h(V,T_h)$ between the black hole thermodynamic pressure $P_h$ and the thermodynamic volume $V$ or the relationship $T_h=T_h(S,P_h)$ between the Hawking temperature $T_h$ and the entropy $S$. Maxwell equal area law states that there is an isobar $P$ or an isotherm $T$ such that $ \text{Area}_{\text{A}}=\text{Area}_{\text{B}}$ in Fig~\ref{fig1}. This description is converted into mathematical language, that is, the expression is in $P_h-V$ plane
\begin{align}
\int_{V_1}^{V_2}P_h dV=P\cdot(V_2-V_1) ~~\Rightarrow~~\int_{V_1}^{V_2}(P-P_h)dV=0.
\label{PV}
\end{align}
or in $T_h-S$ plane
\begin{align}
\int_{S_1}^{S_2}T_h dS=T\cdot(S_2-S_1) ~~\Rightarrow~~\int_{S_1}^{S_2}(T-T_h)dS=0.
\label{TS}
\end{align}

Now, based on the spirit of the definition of generalized free energy in literature~\cite{Xu2024b,Xu2021a,Xu2021b,Wang2024b}, we release the definition of the Maxwell equal area law mentioned above, and introduce generalized free energy $U$ as
\begin{align}
	U^{(P)}=\int(P-P_h)dV, \qquad \text{or} \qquad U^{(T)}=\int(T_h-T)dS,
	\label{U}
\end{align}
which are precisely the indefinite integral expressions of the above equations~\eqref{PV} and~\eqref{TS}. The thermodynamic pressure $P_h$ of the black hole can be written as a function of the thermodynamic volume $V$, and $P$ is an isobar, which is off-shell pressure of the black hole and can assign any positive value in any way. Similarly, the Hawking temperature $T_h$ of the black hole can be written as a function of the thermodynamic entropy $S$, and $T$ is an isotherm, which is off-shell temperature of the black hole and can also assign any positive value in any way. For generalized free energy~\eqref{U}, we can understand it as a canonical ensemble with a pressure $P$. When the ensemble pressure $P$ is equal to the black hole thermodynamic pressure $P_h$, thus the ensemble is entirely composed of real black hole states. It is on-shell condition. Similarly, a canonical ensemble has a temperature $T$ and $T=T_h$ is on-shell condition. For a new function of generalized free energy, what we are currently concerned about is the extremum of this function
\begin{equation}
	\dfrac{dU^{(P)}}{dV}=\dfrac{d}{dV}\left(\int(P-P_h)dV\right)=0~~\Rightarrow~~P=P_h,
\end{equation}
or
\begin{equation}
	\dfrac{dU^{(T)}}{dS}=\dfrac{d}{dS}\left(\int(T_h-T)dS\right)=0~~\Rightarrow~~T=T_h,
\end{equation}
which means that we completely place the real black hole state at the extremum of the generalized free energy. Furthermore, the minimum point of the generalized free energy represents a thermally stable black hole state, while the maximum point represents a thermally unstable state. Of course, for thermally stable states, there is a distinction between stable and metastable states. Local minima represent metastable state, while global minima represent stable state.

It should be noted that for discussing the thermodynamic system of black holes, we generally use the generalized free energy $U=U^{(T)}$, which is more convenient. Indeed, whether it is $U=U^{(T)}$ or $U=U^{(P)}$, although the expression of the generalized free energy obtained is different, the qualitative characteristics are consistent.

Due to thermodynamic fluctuations, black holes show different phase transition behaviors in thus generalized free energy, which can be regarded as a stochastic process. The probability distribution of the evolution of these black hole states (including on-shell states and off-shell states) over time can be described by the Fokker-Planck equation. According to the black hole molecular hypothesis, the phase transition of a black hole is caused by the rearrangement of black hole molecules due to the thermodynamic fluctuation. The Kramers escape rate can be used to obtain information about the stochastic motion of molecules or particles in a potential field~\cite{Risken1988,Zwanzig2001}. If the black hole temperature is much lower than the barrier height, the probability that the molecule is at the lowest point of the well exceeds the probability of reaching the top of the barrier. Even if the molecule reaches the top, it will fall evenly on both sides. If it falls at the lowest point of one potential well, considering thermodynamic fluctuations, it is possible that after staying there for a period of time, it crosses the potential barrier and reaches the lowest point of another potential well.

\begin{figure}[htbp]
	\centering
		\includegraphics[width=90 mm]{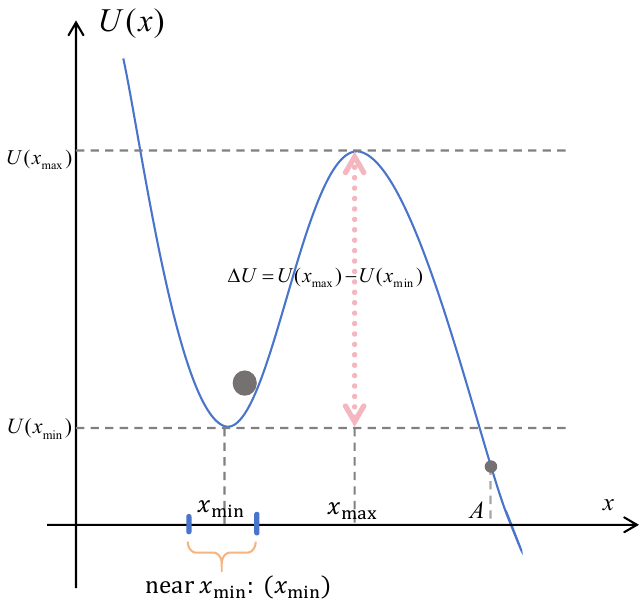}
	\caption{Schematic diagram for calculating Kramers escape rate for potential well.}
	\label{fig12}	
\end{figure}

We start with the Fokker-Planck equation~\cite{Risken1988,Zwanzig2001}
\begin{equation}
\frac{\partial}{\partial t} W(x,t)=-\frac{\partial}{\partial x} J(x,t),
\end{equation}
where the $W(x, t)$ is the probability distribution of black hole molecules and the current $J(x,t)$ is defined as
\begin{equation}
J(x,t)=\dfrac{D e^{U\left(x_{\min }\right)/D} W\left(x_{\min},t\right)}{\int_{x_{\min}}^A e^{U(x)/D} dx},
\label{current}
\end{equation}
where $U(x)$ is the potential or the generalized free energy in our work, and $D$ is the diffusion coefficient, which can be considered constant when the system reaches thermal equilibrium, and we assume that at $x=A$ (A is any position greater than $x_{\max}$), shown in the schematic diagram Fig.~\ref{fig12}, the probability distribution is zero. If we define $p$ as the probability of the particle being inside the well or near $x_{\min}$, then we can find
\begin{equation}
p=W\left(x_{\min},t\right)e^{U(x_{\min})/D}\int_{(x_{\min})} e^{-U(x)/D} dx.
\label{probab}
\end{equation}

The probability $p$ times the Kramers escape rate $r_k$ is just the current $J(x,t)$, hence we can obtain the escape rate~\cite{Risken1988,Zwanzig2001}
\begin{equation}
\frac{1}{r_k}=\frac{p}{J}=\frac{1}{D}\int_{x_{\min}}^A e^{U(x)/D} dx \int_{(x_{\min})} e^{-U(x)/D} dx.
\end{equation}
For above two integrals, we can clearly see that the main contribution of the first integral comes from the regions around $x_{\max}$, while the main contribution of the second integral comes from the regions around $x_{\min}$, shown in Fig.~\ref{fig12}. The Taylor expansions approximation to second order of the potential function $U(x)$ near two extreme points are
\begin{align}
U(x)&\approx U(x_{\max})-\frac12 |U''(x_{\max})|(x-x_{\max})^2,\\
U(x)&\approx U(x_{\min})+\frac12 U''(x_{\min})(x-x_{\min})^2,
\end{align}
and we may extend the above two integrations boundaries to $\pm\infty$, thus the Kramers escape rate can be taken as~\cite{Risken1988,Zwanzig2001}
\begin{equation}
r_k=\frac{\sqrt{\left|U^{\prime \prime}\left(x_{\min }\right) U^{\prime \prime}\left(x_{\max }\right)\right|}}{2 \pi} e^{-\frac{U\left(x_{\max}\right)-U(x_{\min})}{D}}.
\end{equation}

Now we investigate thermodynamic systems of black holes with triple phases, and these three stable states are marked as small black hole state, medium black hole state, and large black hole state, respectively. According to the perspective of generalized free energy, it can be predicted that such a thermodynamic system will have at most five extreme points for its generalized free energy, namely three minima and two maxima, i.e., three potential wells and two potential barriers (it can refer to the diagram (a) in Fig.~\ref{fig2} in subsequent content). We arrange the abscissas of their corresponding extreme points in ascending order, and they are $x_1$, $x_2$, $x_3$, $x_4$, and $x_5$ respectively. For example, we set the Kramers escape rate from the small black hole to the medium black hole as $r_{k1}$, the Kramers escape rate from the medium black hole to the large black hole as $r_{k2}$, the Kramers escape rate from the large black hole to the medium black hole as $r_{k3}$,
\begin{align}\label{rk}
\begin{aligned}
r_{k 1} &=\frac{\sqrt{\left|U^{''}\left(x_1\right) U^{''}\left(x_2\right)\right|}}{2 \pi} e^{-\frac{U\left(x_2\right)-U\left(x_1\right)}{D}},\\
r_{k 2} &=\frac{\sqrt{\left|U^{''}\left(x_3\right) U^{''}\left(x_4\right)\right|}}{2 \pi} e^{-\frac{U\left(x_4\right)-U\left(x_3\right)}{D}},\\
r_{k 3} &=\frac{\sqrt{\left|U^{''}\left(x_5\right) U^{''}\left(x_4\right)\right|}}{2 \pi} e^{-\frac{U\left(x_4\right)-U\left(x_5\right)}{D}}.
\end{aligned}
\end{align}
Without loss of generality, for the numerical calculation of phase transition rate, $D = 10$ will be set for all subsequent calculations.

\section{Phase transition rate between triple phases}
The most familiar thermodynamic system of black holes that can undergo three-phases phase transitions is represented by the Gauss-Bonnet AdS black hole in 6-dimensions and its action is~\cite{Wiltshire1986,Cvetic2002,Cai2002,Cai2013,Wei2022b,Hendi2016a,Hendi2016b}
\begin{equation}
I=\int d^6 x \sqrt{-g}\left[\frac{1}{16 \pi G}\left(\mathcal{R}-2 \Lambda+\alpha \mathcal{L}_{\mathrm{GB}}\right)-\mathcal{L}_{\text {matter}}\right],\label{act}
\end{equation}
where $\Lambda$ is the cosmological constant, and $\alpha$ is the Gauss-Bonnet coupling constant. The Gauss-Bonnet Lagrangian $\mathcal{L}_{\mathrm{GB}}$ and the matter (electromagnetic) Lagrangian $\mathcal{L}_{\mathrm{matter}}$ are in the following form
\begin{equation}
\mathcal{L}_{\mathrm{GB}}=\mathcal{R}_{\mu \nu \gamma \delta} \mathcal{R}^{\mu \nu \gamma \delta}-4 \mathcal{R}_{\mu \nu} \mathcal{R}^{\mu \nu}+\mathcal{R}^2, \qquad
\mathcal{L}_{\text {matter }}=4 \pi \mathcal{F}_{\mu \nu} \mathcal{F}^{\mu \nu}.
\end{equation}
Solving the field equation corresponding to the action (\ref{act}), the static spherically symmetric black hole solution can be obtained
\begin{equation}
ds^2=-f(r) dt^2+f^{-1}(r) dr^2+r^2 d\Omega_4^2,
\end{equation}
where $d\Omega_4^2$ is the line element of the unit $S^4$ and the function $f(r)$ is given by~\cite{Wiltshire1986,Cvetic2002,Cai2002,Cai2013,Wei2022b}
\begin{equation}
f(r)=1+\frac{r^2}{2 \alpha}\left[1-\sqrt{1+4 \alpha\left(\frac{m}{r^5}-\frac{q^2}{r^{8}}-\frac{1}{L^2}\right)}\right],
\end{equation}
where $m$ and $q$ are related to the mass and charge of the black hole, and $L$ is the AdS radius related to the cosmological constant by the equality $\Lambda=-10/L^2$.

The event horizon of the black hole is located at the largest root of $f(r_h)=0$, then the Hawking temperature in terms of event horizon radius $r_h$ is given by~\cite{Wiltshire1986,Cvetic2002,Cai2002,Cai2013,Wei2022b}
\begin{equation}
T_h=\left.\frac{1}{4 \pi} f^{\prime}(r)\right|_{r=r_h}=\frac{3 r_h^2+5 r_h^4 / L^2+\alpha-3 q^2 / r_h^{4}}{4 \pi r_h\left(r_h^2+2 \alpha\right)}.
\end{equation}
Correspondingly, other thermodynamic quantities, including thermodynamic pressure $P_h$, thermodynamic volume $V$, entropy $S$, and charge $Q$ of the 6-dimensional Gauss-Bonnet AdS black hole are~\cite{Wiltshire1986,Cvetic2002,Cai2002,Cai2013,Wei2022b}
\begin{equation}
P_h=\frac{5}{4\pi L^2}, \qquad
V=\frac{8}{15} \pi^2 r_{\mathrm{h}}^5, \qquad S=\frac{2}{3} \pi ^2 r_h^4 \left(\frac{4 \alpha}{r_h^2}+1\right), \qquad Q=\sqrt{6}q.
\end{equation}

Substituting these thermodynamic quantities of the 6-dimensional Gauss-Bonnet AdS black hole into the definition of the generalized free energy (\ref{U}), we have $U=U(r_h)$. For the sake of convenience in description, we replace $r_h$ with $x$, which gives the generalized free energy $U(x)$
 \begin{equation}
U^{(P)}(x)=\frac{1}{3} \pi  \left(-8 \pi \alpha x^2 T_h+2 \alpha x+\frac{8}{5} \pi  P x^5+\frac{Q^2}{3 x^3}-2 \pi  x^4 T_h+2 x^3\right).
\end{equation}
\begin{equation}
U^{(T)}(x)=\frac{1}{3} \pi  \left(-8 \pi \alpha x^2 T+2 \alpha x+\frac{8}{5} \pi  P_h x^5+\frac{Q^2}{3 x^3}-2 \pi  x^4 T+2 x^3\right).\label{gfree}
\end{equation}

\begin{figure}[htb]
	\centering
	\subfigure[coexistence of small, medium, and large black hole states]{
		\includegraphics[width=65 mm]{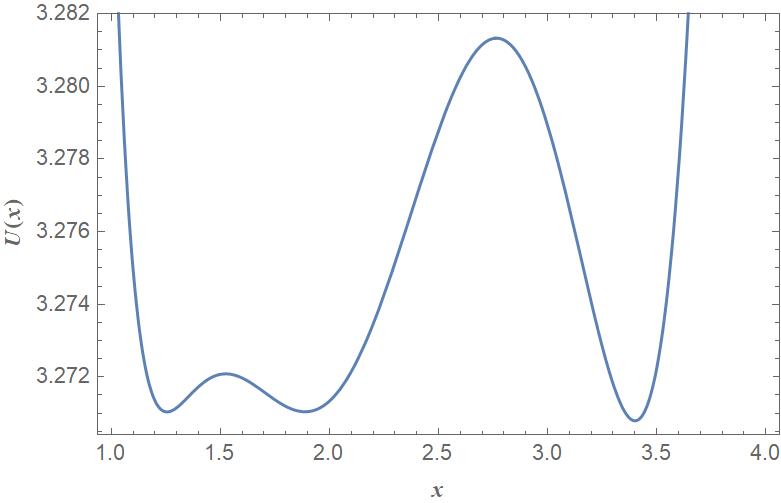} }
	\subfigure[coexistence of small and medium black hole states]{
		\includegraphics[width=65 mm]{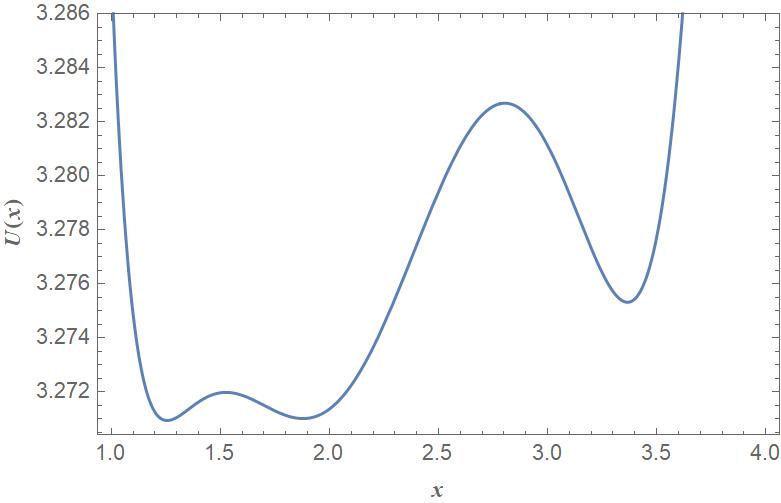} }\\
	\subfigure[coexistence of small and large black hole states]{
		\includegraphics[width=65 mm]{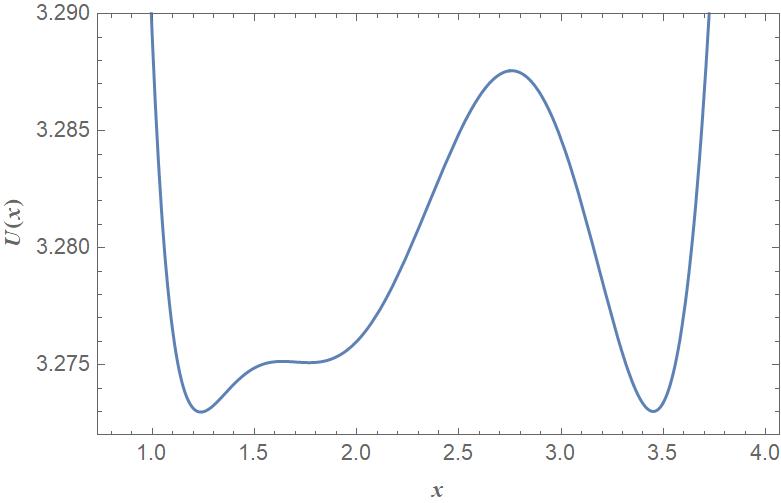} }
		\subfigure[coexistence of medium and large black hole states]{
		\includegraphics[width=65 mm]{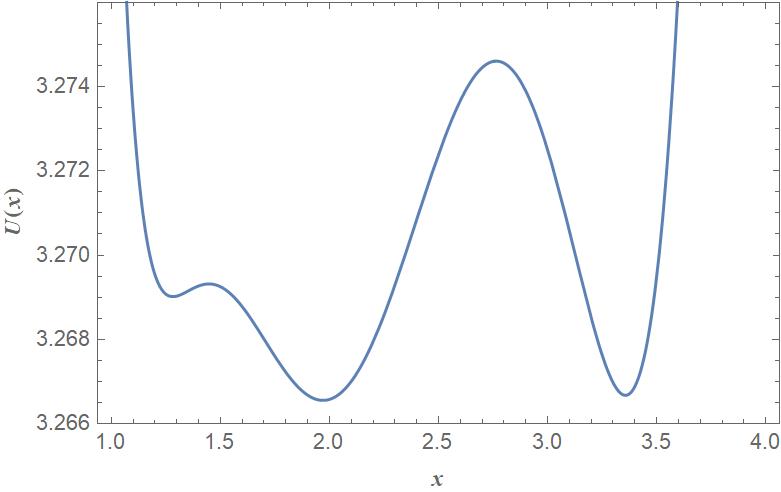} }
	\caption{The behaviors of generalized free energy at $Q=1$ and $\alpha=3.05$ for different temperature $T$ and thermodynamic pressure $P_h$.}\label{fig2}	
\end{figure}

Although the expressions of the generalized free energy obtained are slight different, as mentioned earlier, the qualitative characteristics are consistent, we generally use the generalized free energy $U=U^{(T)}$ to analyze the thermodynamic behavior of black holes. For different temperature $T$ and thermodynamic pressure $P_h$, the generalized free energy has different behaviors. We select representatives of all categories, as shown in the Fig.~\ref{fig2}.

The diagram (a) in Fig.~\ref{fig2} shows the behavior of generalized free energy under three-phases coexistence state under appropriate temperature and thermodynamic pressure. In addition, the thermodynamic system will also exhibit a coexistence of two phases, namely the coexistence of small and medium black hole states in the diagram (b), the coexistence of small and large black hole states in the diagram (c), and the coexistence of medium and large black hole states in the diagram (d). These results indicate that the system will exhibit rich phase transition behavior, which can be divided into three categories: phase transitions between small and medium black hole state, between small and large black hole states, and between medium and large black hole states.

But how do these phase transition processes occur? Which of these three processes is stronger or weaker? When did they start and end respectively? These questions are difficult to obtain clear picture under the usual thermodynamic framework of equilibrium states. Therefore, we rely on Kramers escape rate to provide preliminary answers to these questions, so that we can obtain deeper phase transition information.

\begin{figure}[htb]
	\centering
		\includegraphics[width=90 mm]{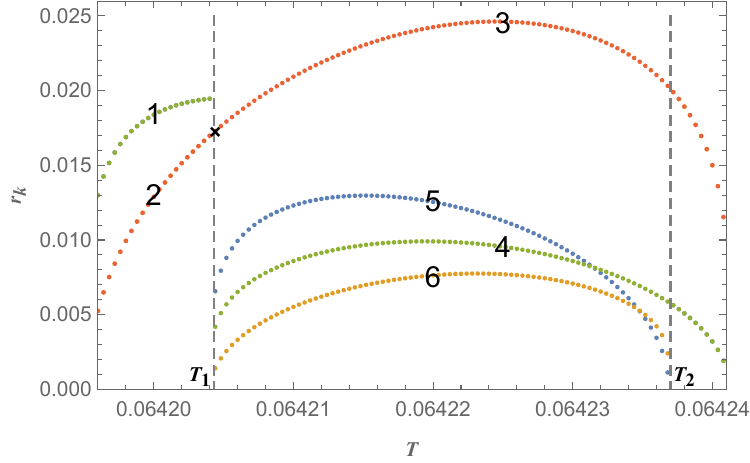}
	\caption{The Kramers escape rate with respect to the temperature at $Q=1$, $\alpha=3.05$ and $P_h=0.0063873$.}
	\label{fig3}	
\end{figure}

Taking advantage of Eqs.~(\ref{rk}) and~(\ref{gfree}), we have obtained the phase transition rate behavior of various processes during the first-order phase transition of the 6-dimensional Gauss-Bonnet AdS black hole, shown in Fig.~\ref{fig3}. From the figure, it can be clearly seen that there are six processes involved in designing the first-order phase transition of a black hole, namely phase transitions between small and medium black hole state, between small and large black hole states, and between medium and large black hole states, and the opposite processes.

\begin{description}
	\item[Process 1] This process represents a transition from a small black hole state to a large black hole state (S $\rightarrow$ L). It can be seen that as the temperature increases, the phase transition rate gradually increases from zero and then stops at the certain rate at $T=T_1$.
	\item[Process 2] This is the reverse process of Process 1, i.e., a transition from a large black hole state to a small black hole state (L $\rightarrow$ S). It also can be seen that as the temperature increases, the phase transition rate gradually increases from zero and then stops at the certain rate at $T=T_1$. Subsequently, this process was continuously linked with Process 3. Overall, the phase transition rate of Process 2 is lower than that of Process 1. Therefore, when the temperature is between 0 and $T_1$, the phase transition of the system is dominated by the transition from a small black hole state to a large black hole state.
	\item[Process 3] This process stands for a transition from a large black hole state to a medium black hole state (L $\rightarrow$ M). We look along the direction of decreasing temperature. As the temperature gradually decreases from a certain value to zero, the large black hole state begins to transition to the medium black hole state. When the temperature decreases to $T_1$, the medium black hole state disappears, and at this point, the large black hole state transitions to the small black hole state until the phase transition ends. Therefore, Processes 2 and 3 are continuously connected.
	\item[Process 4] This process is a transition from a medium black hole state to a large black hole state (M $\rightarrow$ L). When the temperature increases from 0 to $T_1$, the medium black hole state begins to emerge, leading to the transition from the medium black hole state to the large black hole state. Comparing Processes 3 and 4, which are two opposite processes, it can be seen that the system is dominated by Process 3, that is, the transition from the large black hole state to the medium black hole state is dominant.
	
	\item[Process 5] This process represents a transition from a small black hole state to a medium black hole state (S $\rightarrow$ M). It can be seen that this process starts at temperature $T_1$ and ends at temperature $T_2$, showing an overall trend of first increasing and then decreasing. At temperature $T_1$, medium black hole states begin to emerge, while at temperature $T_2$, small black hole states annihilate. At the same time, we notice that Processes 4 and 5 intersect at a specific temperature, indicating that these two processes will form a dynamic equilibrium, that is, the dynamic equilibrium of M $\rightarrow$ L and S $\rightarrow$ M.
	
	\item[Process 6] This is the reverse process of Process 5, i.e., a transition from a medium black hole state to a small black hole state (M $\rightarrow$ S). It also can be seen that this process starts at temperature $T_1$ and ends at temperature $T_2$, showing an overall trend of first increasing and then decreasing. Meanwhile this pair of processes (Processes 5 and 6) intersects at a certain temperature, indicating the formation of a dynamic equilibrium.
\end{description}

\begin{figure}[htbp]
	\centering
		\includegraphics[width=90 mm]{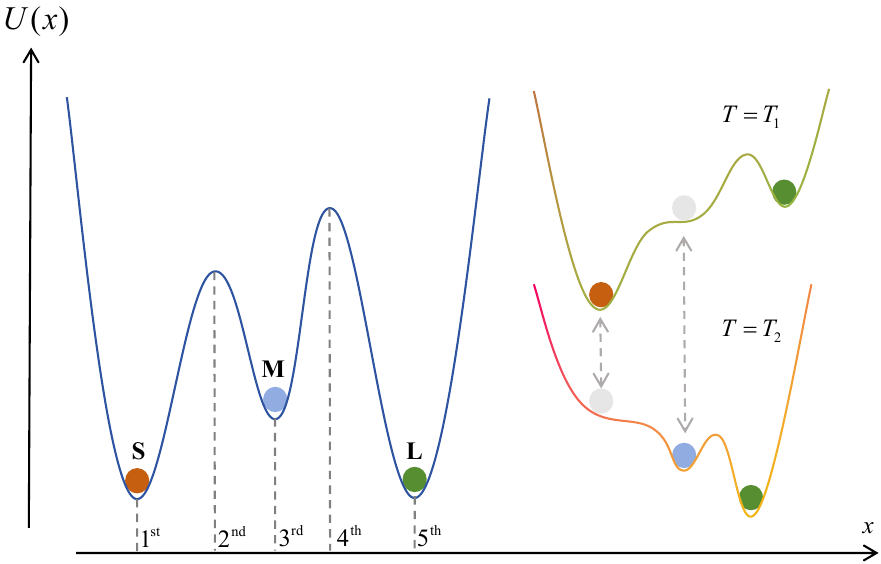}
	\caption{Schematic diagram for two special temperatures $T_1$ and $T_2$ in generalized free energy $U(x)$.}
	\label{fig32}	
\end{figure}

\begin{figure}[htb]
	\centering
		\includegraphics[width=150 mm]{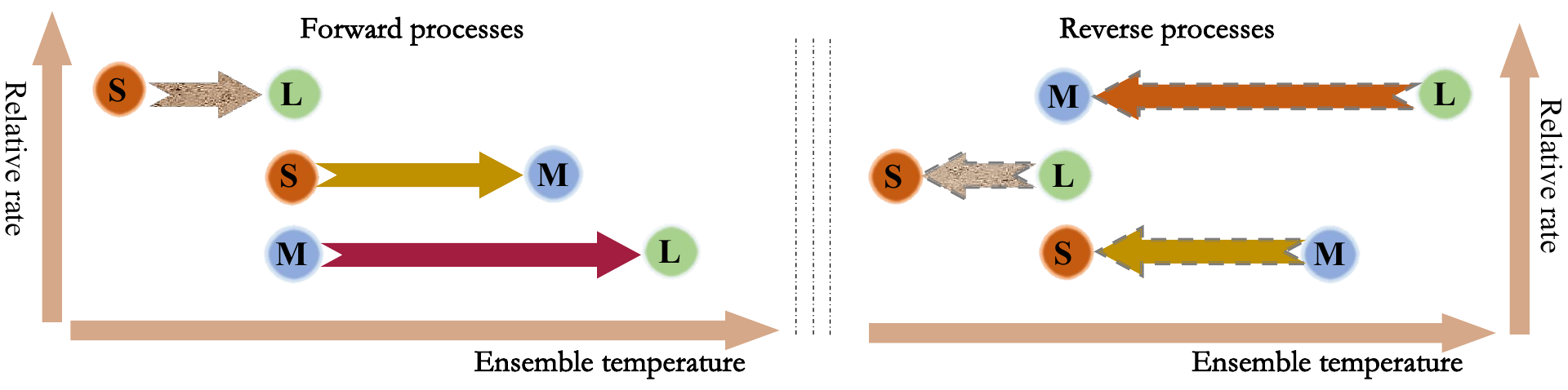}
	\caption{The complete picture of phase transitions of the 6-dimensional Gauss-Bonnet AdS black hole, where ``S'' stands for the small black hole state, ``M'' stands for the medium black hole state and ``L'' stands for the large black hole state}	\label{fig4}	
\end{figure}

From the description of the above six processes, it can be seen that the first-order phase transition behavior of the 6-dimensional Gauss-Bonnet AdS black hole is very rich, and through the analysis of Kramers escape rate, we have also obtained specific information about these different phase transition processes. Overall, it presents three major characteristics:
\begin{itemize}
  \item There are two special temperatures $T_1$ and $T_2$. For generalized free energy $U(x)$, there can be up to five extreme points, see Fig.~\ref{fig32}. At $T_1$, the second and third extremum points of the generalized free energy merge into one (here we arrange the positions of these five extreme points in order according to the direction of $x$ increase), indicating the medium black hole state begins to emerge, forming Processes 3, 4, 5 and 6 while preventing Processes 1 and 2 from proceeding. Mathematically, we require the second and third roots of function $dU(x)/dx=0$ to degenerate into one in order to obtain the value of $T_1$. At $T_2$, it means that the first two extreme points merge into one, indicating the small black hole state begins to disappear, resulting in the termination of Processes 5 and 6. Mathematically, we require the first two positive real roots of function $dU(x)/dx=0$ to degenerate into one to obtain the value of $T_2$.
  \item When the temperature is below $T_1$, the system only has transitions between small and large black hole states; When the temperature is higher than $T_2$, the system only has transitions between medium and large black hole states; When the temperature is between $T_1$ and $T_2$, the phase transition process is complex and rich, with transitions between small, medium, and large black hole states. During these processes, a three-phases coexistence state will be formed. At the same time, two dynamic equilibrium processes will also be formed: the dynamic equilibrium of M $\rightarrow$ L and S $\rightarrow$ M, the dynamic equilibrium of S $\rightarrow$ M and M $\rightarrow$ S.
  \item These six processes can be divided into two categories, namely forward processes (Processes 1, 4 and 5) and reverse processes (Processes 2, 3 and 6), shown in Fig.~\ref{fig4}. For forward processes, there is a dynamic equilibrium of M $\rightarrow$ L and S $\rightarrow$ M. For reverse processes, the L $\rightarrow$ S and L $\rightarrow$ M are continuously connected.
\end{itemize}

\section{Summary}
We already know that the thermodynamic behavior of a 6-dimensional Gauss-Bonnet AdS black hole is similar to that of water, exhibiting three different phases. But we know very little about the process of transitions between these three different phases. In this study, we utilize Kramers escape rate to calculate detailed descriptions of phase transitions in the thermodynamic system of the black hole, enabling us to gain a comprehensive understanding of its thermodynamic properties. We start from the black hole molecular hypothesis and regard the black hole as a thermodynamic system with Brownian motion of molecules. Through the generalized free energy landscape, the phase transition processes between small, medium, and large black hole states are obtained.

For the forward process, that is, the transition from the small black hole state to the large black hole state, the transition from the small black hole state to the medium black hole state, and the transition from the medium black hole state to the large black hole state, as the temperature increases, the transition from the small black hole state to the large black hole state occurs first. At temperature $T_1$, this process terminates and then begins the other two processes. Moreover there is a dynamic equilibrium of the transition from the medium black hole state to the large black hole state and the transition from the small black hole state to the medium black hole state. At temperature $T_2$, the transition from the small black hole state to the medium black hole state terminates. The final process is the transition from the middle black hole state to the large black hole state.

For the reverse process, that is, the transition from the large black hole state to the small black hole state, the transition from the medium black hole state to the small black hole state, and the transition from the large black hole state to the medium black hole state, as the temperature decreases, the transition from the large black hole state to the medium black hole state occurs first. At temperature $T_2$, the transition process from the medium black hole state to the small black hole state begins. As the temperature continues to decrease, there are always two process exist, that is the transitions from the large black hole state to the medium black hole state and from the medium black hole state to the small black hole state. At temperature $T_1$, these two processes terminate, and then begin the transition from a large black hole state to a small black hole state. Moreover, the transition from the large black hole state to the medium black hole state and the transition from the large black hole state to the small black hole state are continuously changing.

This detailed process description provides an overall pciture of the thermodynamic phase transition of the 6-dimensional Gauss-Bonnet AdS black hole, which deepens our understanding of the microscopic thermodynamic behavior of black holes. Moreover, this research approach can be extended to other gravity models to obtain dynamic information on the thermodynamic phase transition of black holes.

\section*{Acknowledgments}
This research is supported by National Natural Science Foundation of China (Grant No. 12105222, No. 12275216, and No. 12247103) ) and supported by the project of Tang Scholar in Northwest University. The authors would like to thank the anonymous referee for the helpful comments that improve this work greatly.

\end{document}